\documentclass[10pt]{iopart}

\usepackage{graphicx}
 \expandafter\let\csname equation*\endcsname\relax
  \expandafter\let\csname endequation*\endcsname\relax
\usepackage{amsmath}
\usepackage{amssymb}
\usepackage{setspace}
\usepackage[numbers]{natbib}
\usepackage{url}

\begin{document}

\title{Vortex dynamics in $\beta$-FeSe single crystals: effects of  proton irradiation and small inhomogeneous stress}

\author{M. L. Amig\'o$^{1,2}$, N. Haberkorn$^{1,2}$, P. P\'erez$^{1,2}$, S. Su\'arez$^{1,2}$ and G. Nieva$^{1,2}$}
\address{$^{1}$ Centro At\'omico Bariloche (CNEA) and Instituto Balseiro (U. N. Cuyo), 8400 Bariloche, R\'{i}o Negro, Argentina.}
\address{$^{2}$ Consejo Nacional de Investigaciones Cient\'{i}ficas y T\'ecnicas, Centro At\'omico Bariloche, Av. Bustillo 9500, 8400 San Carlos de Bariloche, Argentina.}

\ead{amigom@cab.cnea.gov.ar}
\vspace{10pt}
\begin{indented}
\item[]October 2017
\end{indented}

\begin{abstract}

We report on the critical current density $J_c$ and the vortex dynamics of pristine and 3 MeV proton irradiated (cumulative dose equal to 2$\times10^{16}$cm$^{-2}$) $\beta$-FeSe single crystals. 
We also analyze a remarkable dependence of the superconducting critical temperature $T_c$, $J_c$ and the flux creep rate $S$ on the sample mounting method. 
Free-standing crystals present $T_c$=8.4(1)K, which increases to 10.5(1)K when they are fixed to the sample holder by embedding them with GE-7031 varnish. On the other hand, the irradiation has a marginal effect on $T_c$.
The pinning scenario can be ascribed to twin boundaries and random point defects. 
We find that the main effect of irradiation is to increase the density of random point defects, while the embedding mainly reduces the density of twin boundaries.  
Pristine and irradiated crystals present two outstanding features in the temperature dependence of the flux creep rate: 
$S(T)$ presents large values at low temperatures, which can be attributed to small pinning energies, and a plateau at intermediate temperatures, which can be associated with glassy relaxation. From Maley analysis, we observe that the characteristic glassy exponent $\mu$ changes from $\sim$ 1.7 to 1.35-1.4 after proton irradiation.

\end{abstract}



\maketitle
\ioptwocol

\section{Introduction}

The study of the vortex matter in iron-based superconductors (FeBS) is a tool for achieving a better understanding of the interplay between intrinsic superconducting properties, the critical current density ($J_c$) and the vortex depinning mechanisms \cite{SUST0953-2048-25-8-084008}. 
Vortex dynamics in FeBS presents features that can be understood by considering the collective creep theory \cite{PhysRevB.78.224506, RevModPhys.66.1125}.  This theory was originally developed to understand the vortex pinning in superconducting cuprates. 
These materials present a short coherence length, $\xi$, and a large anisotropy, $\gamma=H_{c2}^{ab}/H_{c2}^c$, where $H_{c2}^{ab}$ and $H_{c2}^c$ are the upper superconducting critical fields, for the field applied in the $ab$ plane or along the $c$ axis, respectively.
These characteristics make the pinning energy very small and the resulting relaxation of the persistent critical current higher than what is observed in conventional low temperature superconductors \cite{RevModPhys.66.1125}.

Among FeBS, $\beta$-FeSe presents the simplest crystalline structure since it has no additional structure between the superconducting planes. The electronic properties display signatures of multiband effects\cite{PhysRevLett.115.027006, LouLT27}. The superconducting properties are affected by chemical doping \cite{doi:10.1143/JPSJ.78.074712} and mechanical pressure \cite{PhysRevB.91.174510, doi:10.1063/1.4903922}. In addition, a slight increment of the superconducting transition temperature $T_c$ in electron irradiated samples was reported \cite{PhysRevB.94.064521}. 
An important characteristic of $\beta$-FeSe is the tetragonal to orthorhombic structural transformation at $T_s \sim 90\,$K \cite{PhysRevLett.103.057002}. This structural transition produces twin boundary (TBs) planes, which are aligned at $\sim$ 45$^{\circ}$ of the $ a $ and $b$ directions \cite{PhysRevB.79.180508, PhysRevLett.109.137004}. 
The presence of TBs in $\beta$-FeSe may provide a non negligible contribution to the vortex pinning \cite{PhysRevLett.109.137004, nature}. 
However, its contribution to the flux creep mechanism has not been previously discussed. According to reference \cite{PhysRevB.92.144509}, the vortex dynamics in $\beta$-FeSe single crystals is governed by a combination of random disorder assisted by a small density of nanometric defects. 

The strength of the pinning potential depends on the intrinsic superconducting parameters and on the type of pinning centers \cite{RevModPhys.66.1125}. 
$\beta$-FeSe single crystals present $T_c = 8.4(1)\,$K, penetration depth $\lambda_{ab}(0) \sim 445\,$ nm\cite{PhysRevB.88.174512}, coherence length $\xi_{ab}  (0) \sim 4.4\,$nm and a temperature dependent anisotropy, which satisfies $\gamma(T \rightarrow T_c) \sim 3$ and $\gamma(T\rightarrow 0)\sim1$ \cite{Amigó2015}.
The intrinsic thermal fluctuations can be parameterized by the Ginzburg number, $G_i=\frac{1}{2 }(\gamma T_c / H_c^2 \xi^3)^2$, which measures the relative size of the minimal ($T=0$) condensation energy $H_c^2(0)\xi^3(0) / \gamma$  within a coherence volume \cite{RevModPhys.66.1125}.  Here $H_c(0)=\phi_0 / 2  \sqrt{2} \lambda(0)\xi(0)  \sim$ 1.2\,kOe is the thermodynamic critical field, where $\phi_0$ is the magnetic flux quantum. 
For $\beta$-FeSe, $G_i \sim 5\times10^{-5}$ and the theoretical depairing critical current density  $J_0(T=0)=c H_c / 3 \sqrt{6} \pi \lambda \approx $11.3 MAcm$^{-2}$, where $c$ is the speed of the light. This value of $G_i$ is between those of low temperature superconductors ($\sim 10^{-8}$) and cuprates ($\sim 10^{-2}$).

In this work we report on the superconducting properties, the critical current density and the vortex dynamics of pristine and of 3MeV proton irradiated $\beta$-FeSe single crystals based on magnetic and electrical transport measurements.
 $T_c$ is unaffected by irradiation but is affected by the sample mounting method. Free-standing single crystals present $T_c=8.4(1)$\,K, which is increased to 10.5(1)\,K when the sample is fixed to the sample holder with GE-7031 (polyvinyl phenolic non-magnetic varnish).  This change can be attributed to the stress produced by differential thermal contraction \cite{Morelock:ks5362}.
We observe that both the dependence on temperature, $T$, and on magnetic field, $H$, of the critical current density, $J_c (H, T)$, and of the flux creep rate, $S(H, T)$, are affected by the sample mounting method and by the irradiation. 
The results are analyzed considering the collective creep theory. The characteristic glassy exponents $\mu$ are obtained by using Maley analysis \cite{PhysRevB.42.2639}. The results show that at intermediate temperatures the vortex relaxation in pristine samples presents a glassy exponent $\mu$ of $\sim$ 1.7, which is reduced to 1.35 -- 1.4 after proton irradiation.

\section{Methods}

The $\beta$-FeSe single crystals were grown inside a sealed quartz ampule using $\tfrac{1}{3}$KCl:$\tfrac{2}{3}$AlCl$_{3}$ flux in a temperature gradient of about 5$^\circ$C/cm with the hotter end of the ampule at 395\,$^\circ$C for 45 days \cite{LouLT27}. 
The phase purity of each crystal was verified by X-ray diffraction (XRD) using a PANalytical Empyrean equipment with Ni filtered Cu K$_\alpha$ radiation.  

\begin{figure}[h]
\begin{center}
\includegraphics[width=0.5\textwidth]{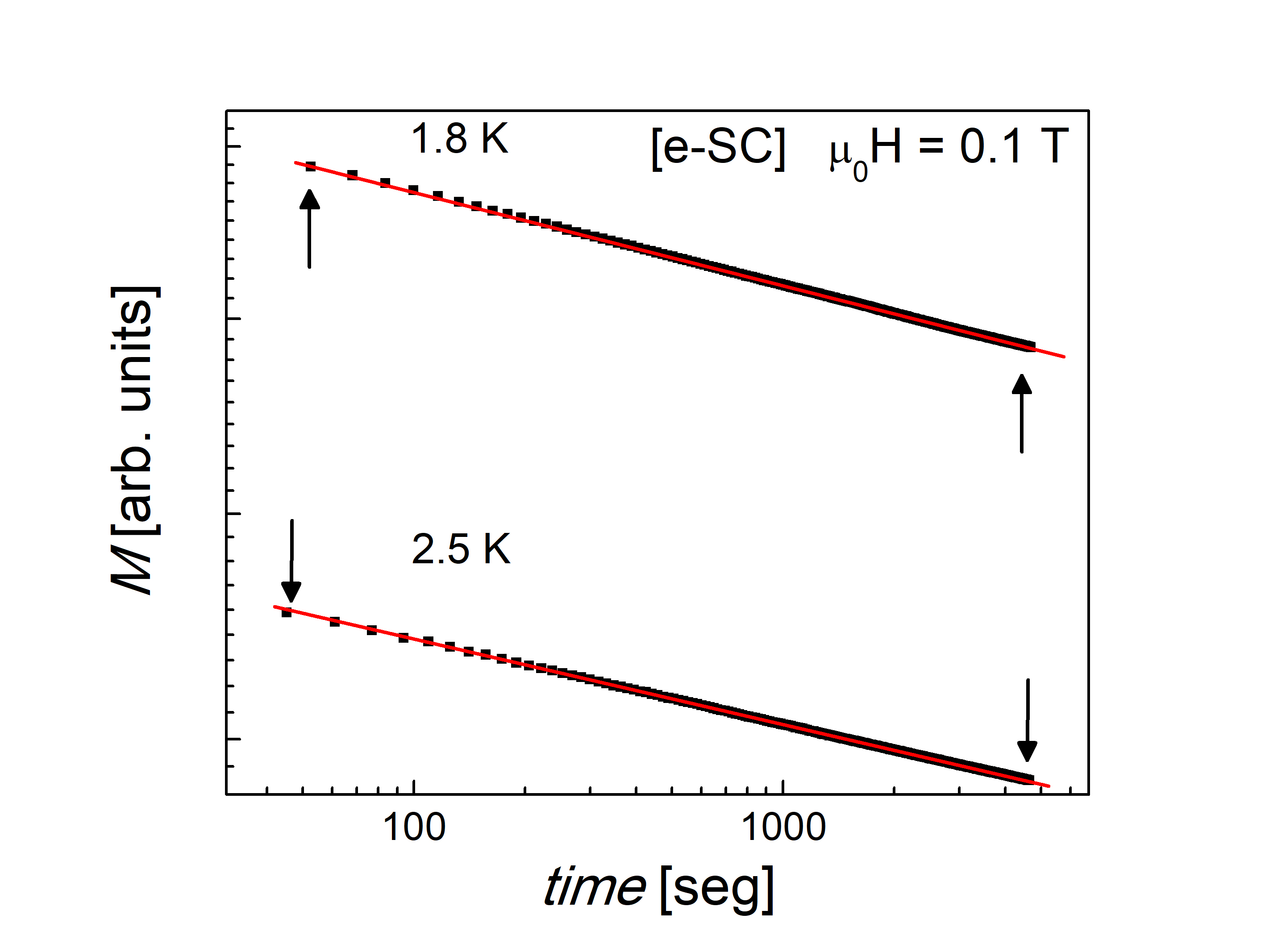}
\end{center}
\caption{Magnetization as a function of time for $T$=1.8 and 2.5\,K and an applied magnetic field of $\mu_0 H=0.1$\,T for a crystal embedded in GE-7031 varnish. The arrows show the range of time taken into account to obtain the flux creep rate.}\label{Creep}
\end{figure}

Measurements of the magnetization, $M$, were performed using a superconducting quantum interference device (SQUID) magnetometer with the magnetic field parallel to the $c$ axis ($H \parallel c$). $J_c$ was estimated by applying the Bean critical-state model to the hysteresis loop. According to this model,  $J_c=20\Delta M/(d w^2 (l-w/3))$, where $\Delta M$ is the difference in magnetization between the top and bottom branches of the hysteresis loop, and  $ d$, $ w$, and $l$ are the thickness, width, and length of the sample $(l > w)$, respectively. 
The flux creep rate, $ S=-\frac{\text{d}ln(J)}{\text{d}ln(t)}$ was recorded as a function of time, $t$, over periods of one hour.
The magnetization of the sample holder was measured and subtracted from the data by averaging the initial points of the time relaxation for the lower and upper magnetic branches. The initial time was adjusted considering the best correlation factor in the log-log fitting of the $J_c (t)$ dependence. Figure \ref{Creep} presents a typical example of the magnetization as a function of time for $T$=1.8 and 2.5\,K and an applied magnetic field of $\mu_0 H=0.1$\,T. The arrows show the range of time that was taken into account to obtain the flux creep rate. The initial critical state for each creep measurement was generated by applying a field $H \sim 4 H^{\ast}$, where $ H^{\ast}$ is the field for the full-flux penetration \cite{RevModPhys.68.911}. The data shown in this paper were obtained with a scan length of 3\,cm. 

For magnetic measurements, two different sample mounting procedures were used:
 (i) the crystal was held free-standing on a Delrin disk (sample holder) and covered with Teflon tape to avoid movement due to magnetic  torque, (ii) the single crystal was fixed to the Delrin disk using GE-7031 varnish dried at room temperature. Between measurements, the varnish was dissolved and rinsed away with 50:50 v/v toluene - isopropil  alcohol  mixture.  The measurements were reproducible between successive mounting changes. 
In particular, no difference in the superconducting critical temperature was observed in the successive heating cycles for each mounting configuration. 
The differential thermal expansion between the single crystals of $\beta$-FeSe and the GE-7031 varnish is expected to create a stress on the embedded sample. Considering the thermal expansion coefficients, the elastic moduli and the Poisson ratio, we estimate that, at low temperatures, a positive strain ($\approx$0.55\,GPa) is applied on the embedded sample \cite{Morelock:ks5362}. 
In addition, it is expected that the inhomogeneous stress produced by the GE-7031 varnish affect the density of TBs that appear during the structural transition at $T_s$ \cite{PhysRevLett.109.137004}. 
It is important to note that other sample mounting methods, such as holding the crystal with vacuum grease, also have an effect on the measurements. The changes in the superconducting critical temperature with different sample mounting methods are summarized in Table \ref{Tc}. We find that for the GE-7031 varnish the effect is larger, and therefore, we analyze in detail this case. Furthermore, in the literature there are some reports on the effects of sample mounting induced strain \cite{doi:10.1063/1.4903922, WANG20171}. Consequently, in the case of $\beta$-FeSe and FeSe$_\text{1-x}$Te$_\text{x}$ a word of caution is in order on the choice of sample mounting method due to the consequences on the physical properties measured.
\begin{table}[t]
\caption{\label{Tc} $T_{c}$, in Kelvin, measured for different sample mounting method of pristine single crystals.}
\centering  \scalebox{1}{
\begin{tabular}[b]{l c c c c }
\hline \hline
&      free-standing  &    vacuum grease &  GE-7031\\ 
&					&					& varnish  \\
\hline \hline
   $T_c$[K]  &    8.4(1)   &   9.9(5)  & 10.5(1) \\
  \hline \hline
\end{tabular}}
\end{table}

To measure the electrical resistance, $R$, a conventional four wire method was used. The samples were placed on a sapphire sample holder, and again, we considered both the case in which the sample is free-standing or is embedded in a GE-7031 dried drop. 
In the first case, the thermal contact to the sapphire holder is provided by the gold wires attached with silver paint. 

In a first stage, the measurements were done in a pristine crystal. Then the measurements were repeated in the single crystal irradiated with 3MeV proton with a cumulative dose of $2\times10^{16}$ cm$^{-2}$. Irradiation with 3 MeV protons produces mostly Frenkel pairs, i.e. random point defects. This dose was chosen because it is known to improve the pinning in cuprates \cite{PhysRevLett.65.1164} and FeBS \cite{PhysRevB.85.014522, 0953-2048-27-9-095004}. 

The studied single crystals initially had the following dimensions: $d$ = 0.068 mm, $w$ = 0.61 mm and $l$ = 0.68 mm. After proton irradiation the single crystal was cleaved and $d$ decreased to 0.055 mm. 
In this work, we use the following notation: 
[f-SC] corresponds to free-standing pristine single crystal,
[e-SC] to pristine single crystal embedded in GE-7031  varnish,
[f-SC-irr] to free-standing proton irradiated single crystal,
and [e-SC-irr] to irradiated single crystal embedded in GE-7031 varnish.

\begin{figure}[h]
\begin{center}
\includegraphics[width=0.46\textwidth]{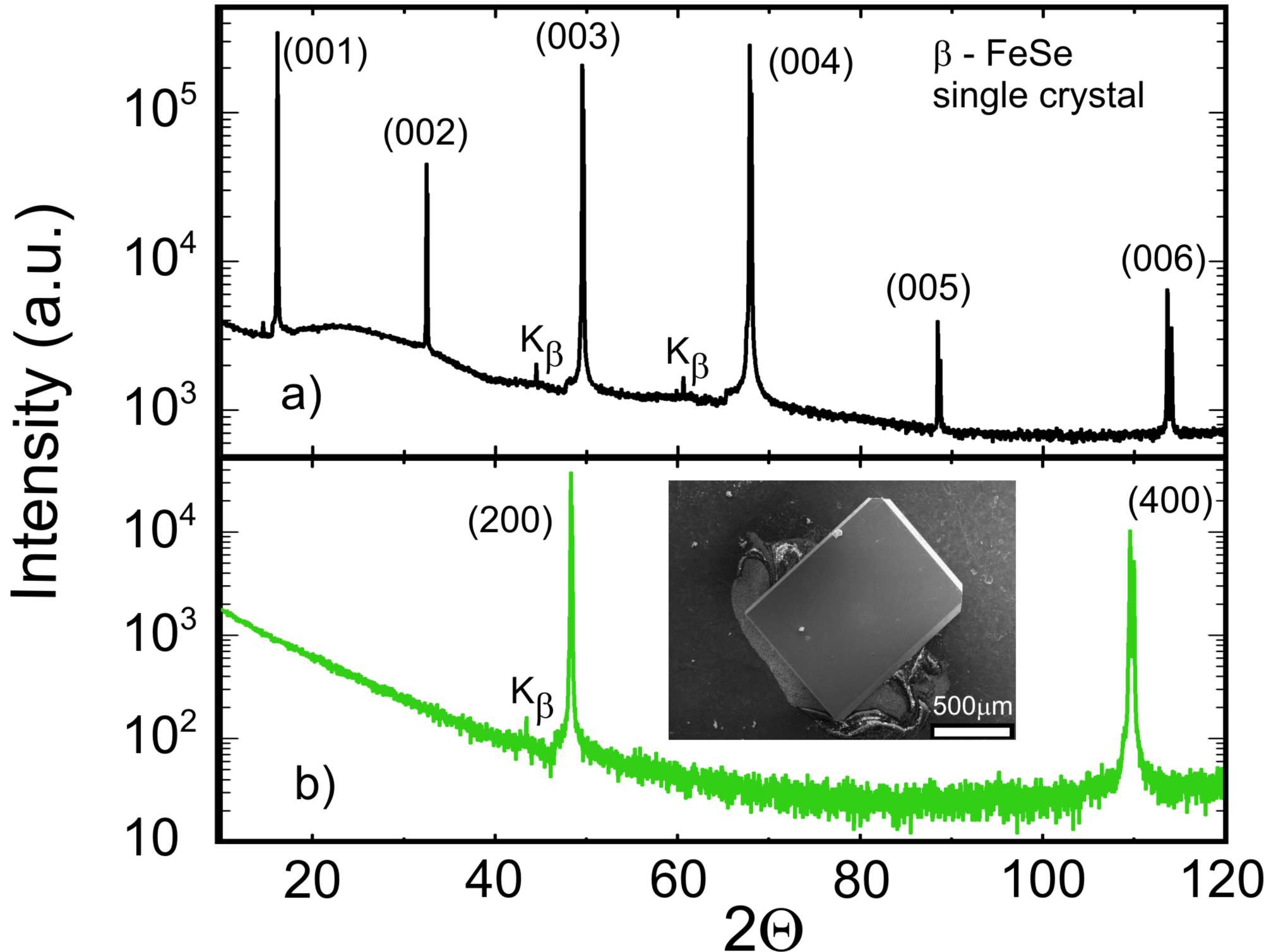}
\end{center}
\caption{ Room temperature X-ray diffraction patterns of a typical $\beta$-FeSe single crystal showing the  $ c$-axis a) and the $ a$-axis b) reflections. The background corresponds to the amorphous sample holder signal. The peaks corresponding to the remaining Cu K$_{\beta}$ radiation are explicitly labelled. $Inset$: Scanning electron microscope image of a typical single crystal.}\label{RX}
\end{figure}

\section{Results and Discussion} 
\subsection{Crystalline structure}

Fig. \ref{RX} shows a typical XRD pattern obtained along the $(00l)$ and the $(h00)$ directions.
The single crystals present a tetragonal P4/nmm (129) unit cell, with lattice parameters $a = b =$ 0.377(1)\,nm  and $c =$ 0.552(1)\,nm. These values are in agreement with those reported in Ref. \cite{C2CE26857D}.
The inset presents a scanning electron microscope image of a typical $\beta$-FeSe single crystal. All the single crystals exhibit a platelet-like morphology with the $c$ axis perpendicular to the plane of the plate.

\subsection{Superconducting transition temperature}

\begin{figure}[t]
\begin{center}
\includegraphics[width=0.46\textwidth]{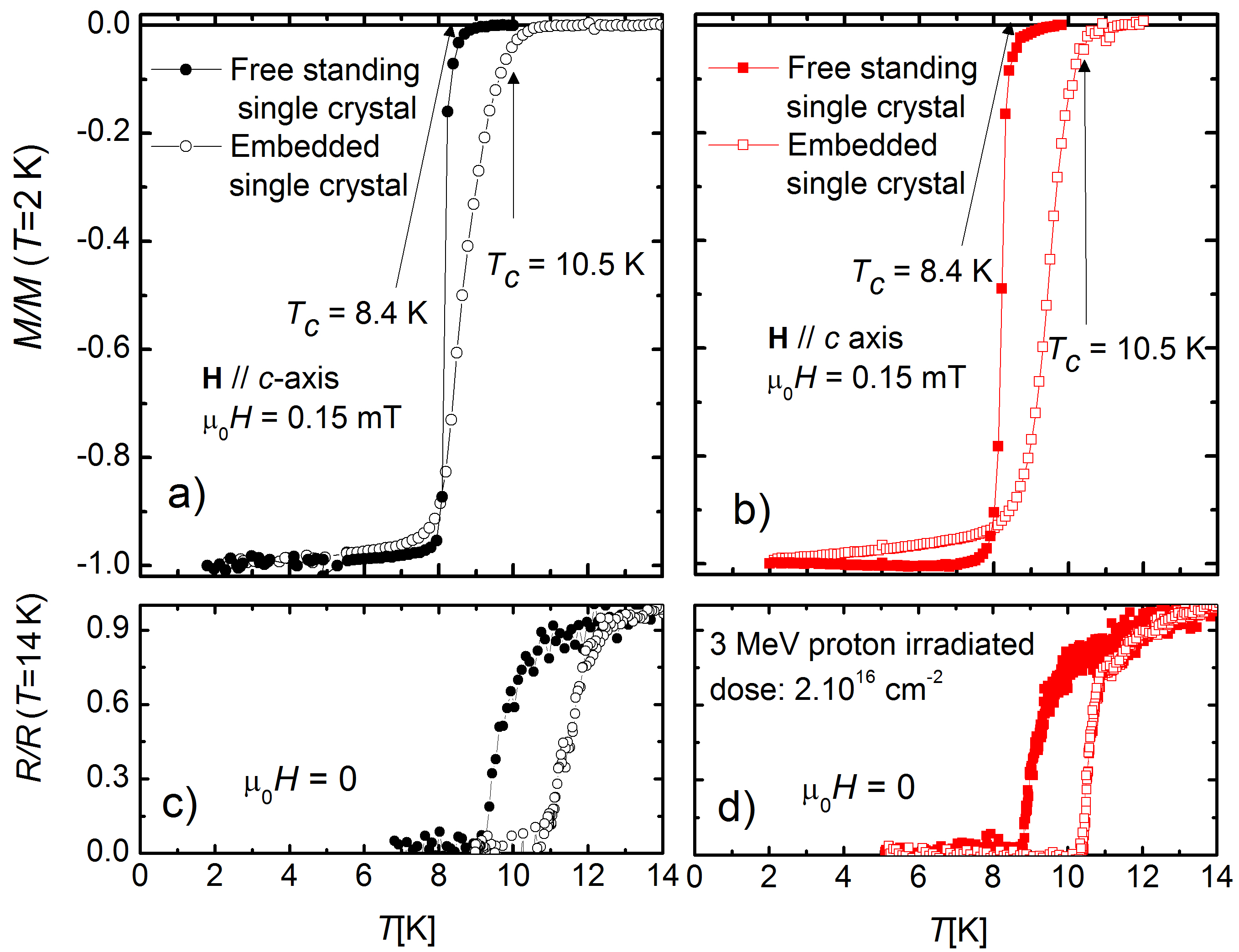}
\end{center}
\caption{Transition temperature measurements for $\beta$-FeSe single crystals. The temperature dependence of the magnetization is shown for the pristine a) and irradiated single crystal b) for the free-standing (solid symbols) and  embedded in GE-7031 varnish (open symbols) sample mounting configurations. A magnetic field of $\mu_0 H=0.15$\,mT  parallel to the $c$ axis applied after zero field cooling was used. The magnetization in each case was normalized by its value at 2\,K. 
The resistance at zero magnetic field and normalized at 14\,K is shown for the pristine sample in c) and for the irradiated in d) both for free-standing (solid symbols) and embedded in GE-7031 varnish (open symbols) mounting configurations. }\label{Tcs}
\end{figure}

Fig. \ref{Tcs}a presents the temperature dependence of the normalized magnetization, $M/M(T=2\,\text{K}$), obtained using a magnetic field $\mu_0 H = 0.15\,$mT after a zero field cooling (ZFC), for the samples [f-SC] and  [e-SC]. Fig. \ref{Tcs}b shows similar data for [f-SC-irr] and [e-SC-irr]. 
Both free-standing ([f-SC]  and [f-SC-irr]) samples have $T_c = 8.4(1)$\,K as measured from the transition onset. For the embedded samples ([e-SC]  and [e-SC-irr]) $T_c$ increases to $10.5(1)$\,K. This increment is similar to that obtained applying a hydrostatic pressure of $\sim  0.3$ GPa\cite{PhysRevB.91.174510}.
In addition, in the embedded samples, the wider transition suggests the existence of inhomogeneous strains induced by the sample mounting method.
Figs. \ref{Tcs}c and \ref{Tcs}d show the temperature dependence of the normalized resistance, $R/R(T=14\,\text{K})$, measured at zero magnetic field. 
The zero resistance temperature agrees with the onset of the transition in $M (T)$. This indicates that a percolation superconducting path across the sample occurs simultaneously with the loss of the screening in zero field cooling measurements. 

\subsection{Critical currents densities and vortex relaxation mechanism}

\begin{figure}[b]
\begin{center}
\includegraphics[width=0.46\textwidth]{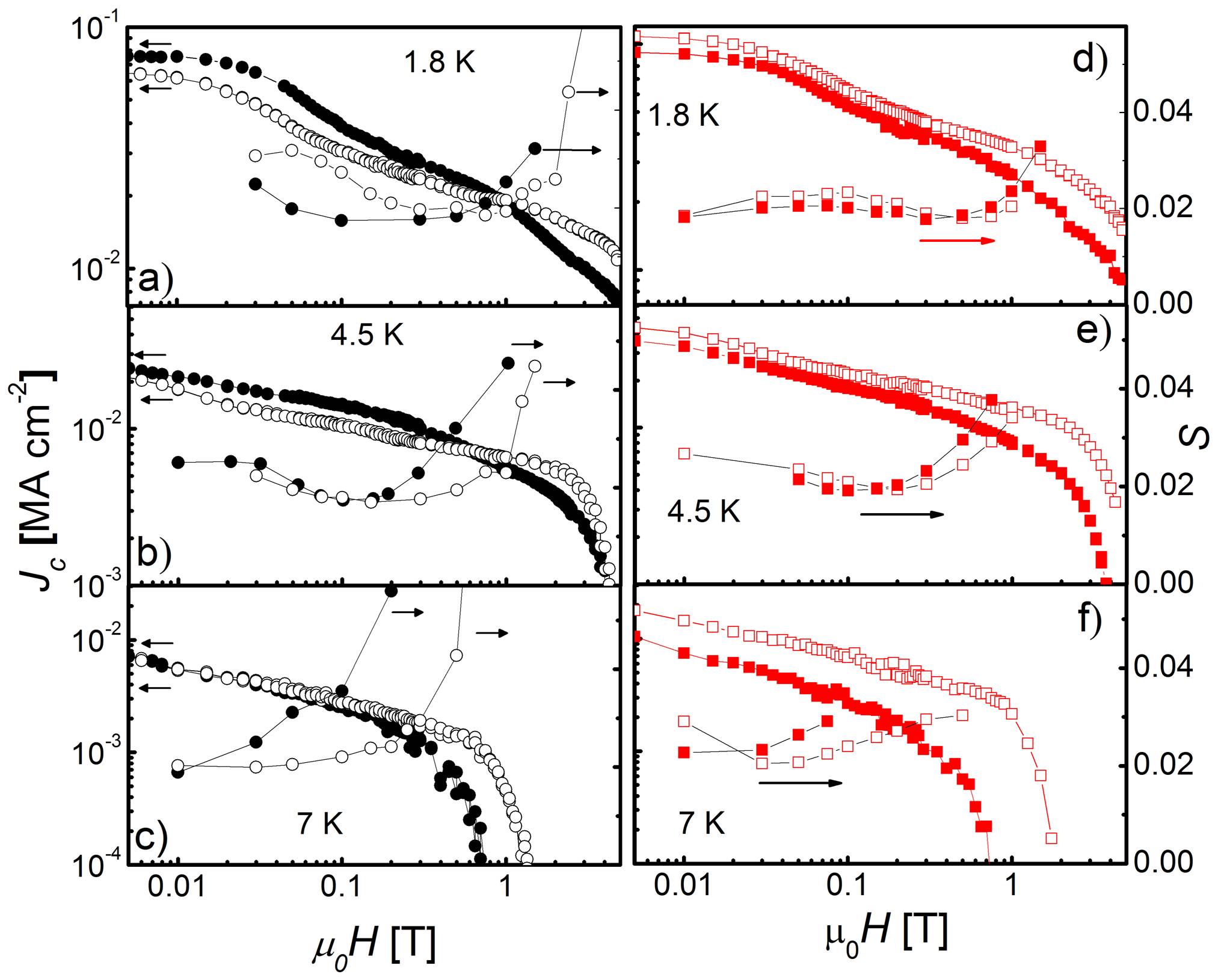}
\end{center}
\caption{Magnetic field dependence of the critical current densities $(J_c)$ and flux creep rates $(S)$ at a) 1.8\,K, b) 4.5\,K, and c) 7\,K in [f-SC] (solid symbols) and [e-SC] (open symbols). The corresponding $J_c$ and $S$ for the 3Mev proton irradiated sample are shown in  d), e) and f) panels. In all cases, the applied magnetic field is parallel to the $c$ axis of the crystal.}\label{JyS}
\end{figure}

Figs. \ref{JyS}a, \ref{JyS}b and \ref{JyS}c show $J_c (H)$ (left axis) and $S(H)$ (right axis) at temperatures 1.8\,K, 4.5\,K and 7\,K for [f-SC] and [e-SC].
Both $J_c (H)$ and $S(H)$ display a modulation with magnetic field usually attributed to changes in the vortex bundle size \cite{PhysRevB.50.7188, PhysRevB.81.174517}. 
At $T=1.8\,$K, the self-field critical current density, $J_{csf}$, is $\sim$ 0.08 MA.cm$^{-2}$ for [f-SC] and  $\sim$ 0.066MA.cm$^{-2}$ for [e-SC]. 
The low ratio $J_{csf} /J_0\sim$0.06\% is inside of the predictions for weak pinning produced mainly by random point defects
(size smaller than $\xi$) \cite{RevModPhys.66.1125}.
It is noticeable that the sample with larger $T_c$ presents a smaller value of $J_{csf}$. This also occurs for other temperatures $T \lesssim T_c/2$ suggesting that the pinning at low fields is originated by a different type of defects in free-standing or in varnished embedded samples. 
The main structural difference between the free-standing and the embedded sample can be related to the density of TBs originated during the structural transition at $T_s$. The inhomogeneous stress associated with the embedding configuration is expected to reduce the density of TBs.
In this scenario, the vortex pinning landscape of [f-SC] is originated by TBs and random point defects, whereas a smaller contribution of TBs to the pinning is expected in [e-SC]. 
A low density of TBs is expected to enhance the pinning mainly at low fields, whereas a smaller contribution is expected at intermediate fields. 
This scenario is consistent with the $S(H)$ dependence observed at 1.8 K. At low fields ($\mu_0H < 0.3$ T), [f-SC] displays smaller $S$ values than [e-SC], but both mounting configurations display similar values at intermediate fields (0.3 T -- 1 T). At high fields, independently of the sample holding method a crossover to fast creep ($S$ is strongly increased) is observed. This is usually associated with an elastic to plastic crossover in the vortex relaxation \cite{PhysRevB.78.224506}. 

Figs. \ref{JyS}d, \ref{JyS}e and \ref{JyS}f show $J_c (H)$ (left axis) and $S(H)$ (right axis) at 1.8\,K, 4.5\,K and 7\,K for [f-SC-irr] and [e-SC-irr]. Both sample mountings display similar $J_c(H)$ and $S(H)$ dependences, which indicates that the pinning in irradiated samples is dominated by the same mechanisms. 
The disappearance in [e-SC-irr] of the peak observed in [e-SC] in $S(H)$ at 1.8\,K and low fields suggests that the vortex pinning mechanism is changed after irradiation. This fact could be associated with the presence of a high density of random point defects and some small nanoclusters with a size larger than $\xi$ (strong pinning centers) \cite{doi:10.1063/1.4821440}.

\begin{table}[b]
\caption{\label{Hc2} $H_{c2}$, in Tesla, measured from transport experiments for the samples [f-SC], [e-SC], [f-SC-irr] and [e-SC-irr].\\   * extrapolated values}
\centering  \scalebox{1}{
\begin{tabular}[b]{l c c c c c c }
\hline \hline
&     &   [f-SC]   &   [e-SC] & [f-SC-irr] & [e-SC-irr]  \\
\hline \hline
 1.8\,K   &   &   13.6   &   18*  &  14.4   &  17.7*   \\
 4.5\,K  &   &   8.3   &   12.3 &  9.2  & 12.3   \\
 7\,K  &   &   3.7    &   7.4 &  4.8  & 7.3  \\
  \hline \hline
\end{tabular}}
\end{table}

\begin{figure}[t]
\begin{center}
\includegraphics[width=0.46\textwidth]{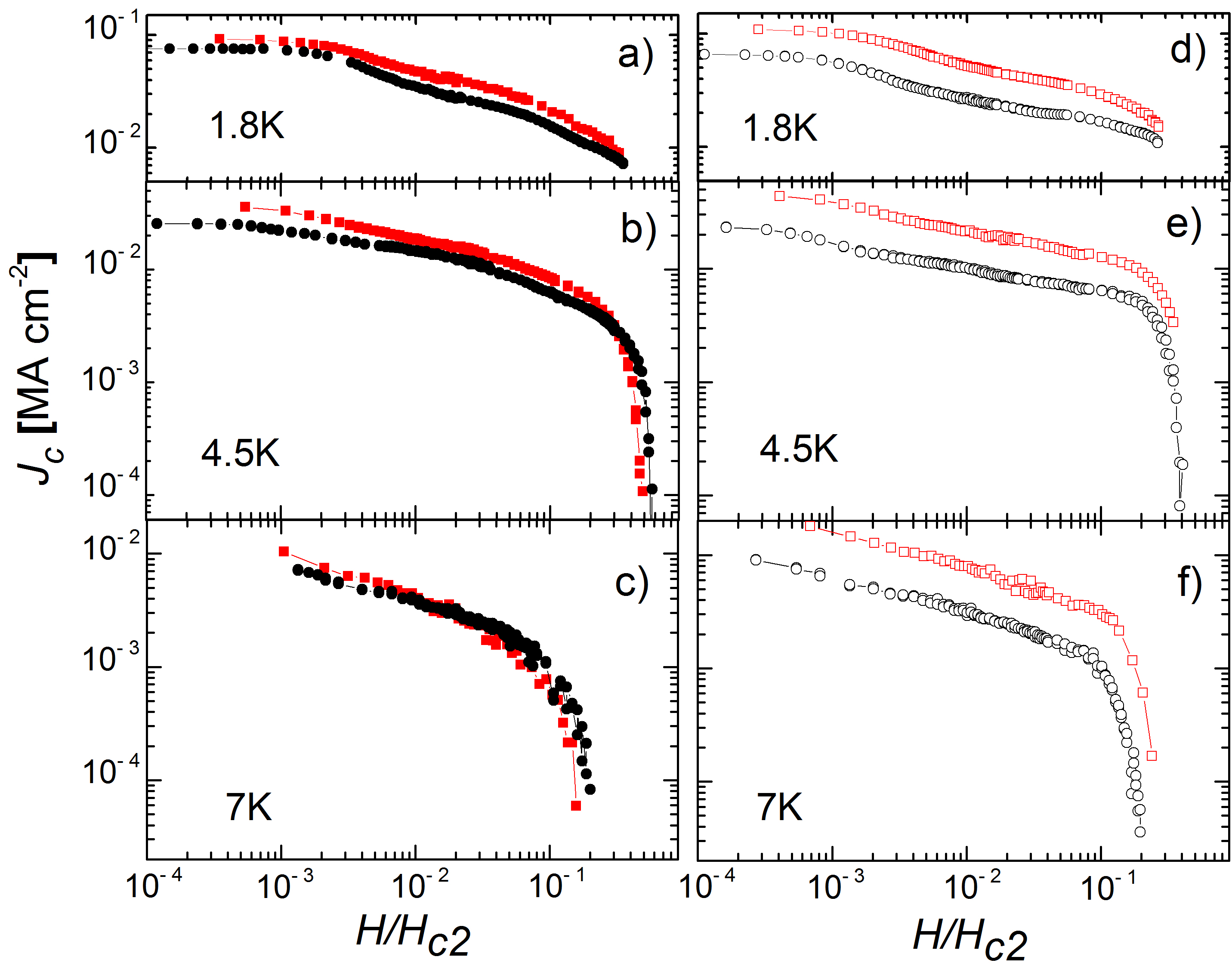}
\end{center}
\caption{Critical current densities at a) 1.8\,K, b) 4.5\,K, and c) 7\,K in [f-SC] (circular black symbols) and [f-SC-irr] (squared red symbols)  as function of magnetic field normalized by the corresponding $H_{c2}$. $J_c$ values for the embedded sample are shown in  d), e) and f) panels also as a function of the field normalized by $H_{c2}$.}\label{J comp}
\end{figure}

To make a proper analysis of the effects of the irradiation on $J_c(H)$ and $S(H)$, we have measured with transport properties the values of $H_{c2}$, defined as the onset of the transition. The results of $H_{c2}$ for the different cases are presented in Table \ref{Hc2}. In the following, we analyze both the critical current density and the flux creep rate as function of $H/H_{c2}$.

\begin{figure}[b]
\begin{center}
\includegraphics[width=0.46\textwidth]{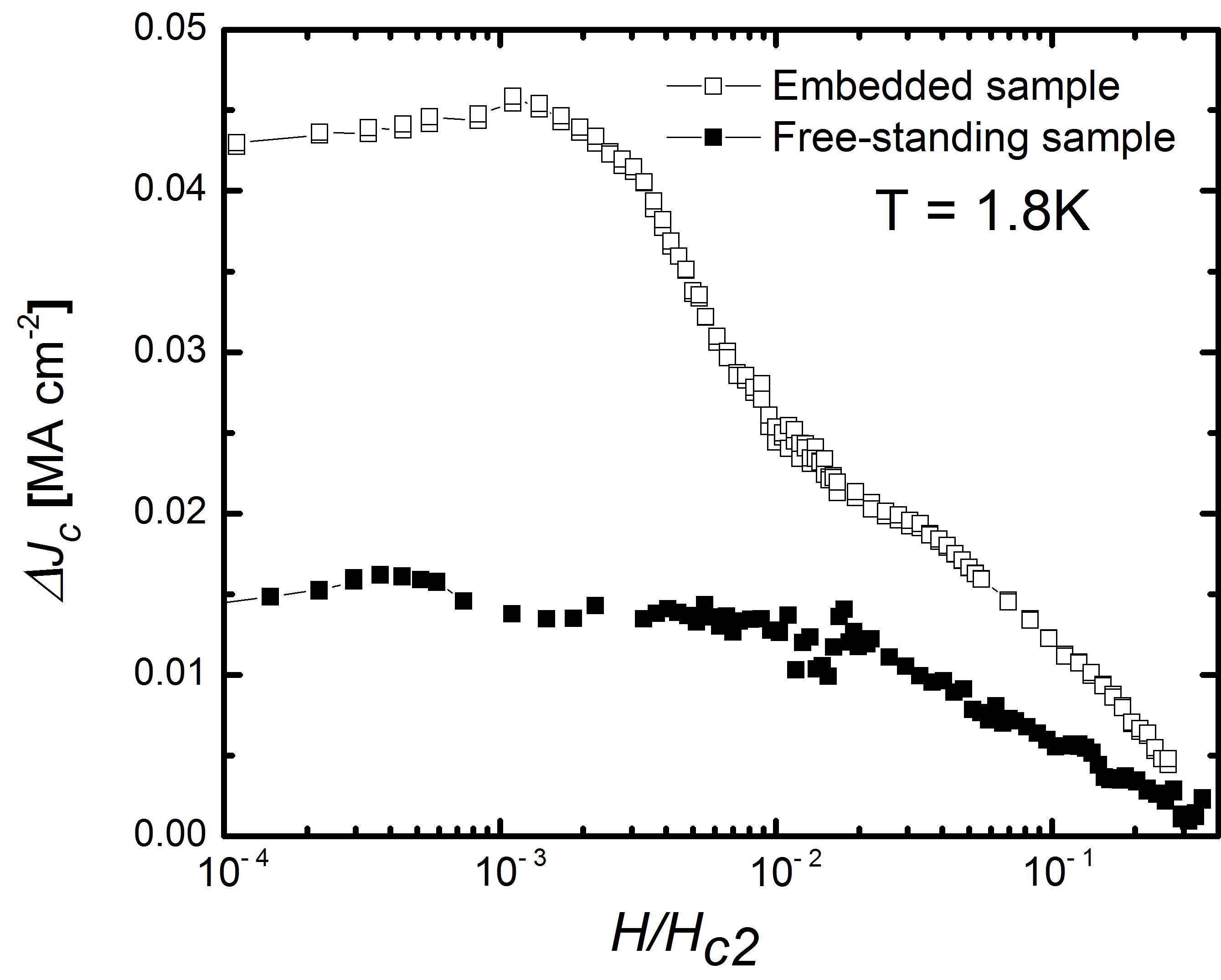}
\end{center}
\caption{Difference in the critical current densities $(J_c)$ at 1.8\,K  before and after proton irradiation for the sample free-standing (full symbols) and embedded in GE-7031 varnish (open symbols).}\label{deltaJ}
\end{figure}

In Fig. \ref{J comp} we compare  $J_c(H/H_{c2})$ for the sample before and after irradiation using the same data shown in Fig. \ref{JyS}. In panels a) 1.8\,K, b) 4.5\,K and c) 7\,K for the free standing sample, while in panels  d), e) and f) the same comparison is made for the embedded sample case.
In general, the proton irradiation produces an increase in the $J_c(H, T)$, as expected. An exception is observed for free-standing samples at 7\,K, which can be attributed to an increment in the vortex fluctuations close to $T_c$ produced by the irradiation damage.
The enhancement of  $J_c$ is very important in the case of the embedded sample even at the higher measured temperature. To examine in more detail the influence of the irradiation on the $J_c(H)$ dependences we analyzed the difference $\Delta J_c(H/H_{c2}) = J_c^{\text{[i-SC-irr]}}-J_c^{\text{[i-SC]}}$ where i= e or f at 1.8\,K. 
Fig. \ref{deltaJ} shows the results obtained for both mounting configurations, which are quantitatively different.
To understand the differences, it is useful to consider the pinning landscape for each sample.
The inclusion of additional random disorder and nanoclusters by irradiation should affect significantly the pinning above the matching field produced by TBs \cite{RevModPhys.66.1125}. 
This is consistent with the fact that when lowering $H/H_{c2}$, $\Delta J_c$ saturates for the [f-SC] while it presents an additional increase at low fields for the [e-SC] (which presumably has a lower density of TBs).

\begin{figure}[t]
\begin{center}
\includegraphics[width=0.46\textwidth]{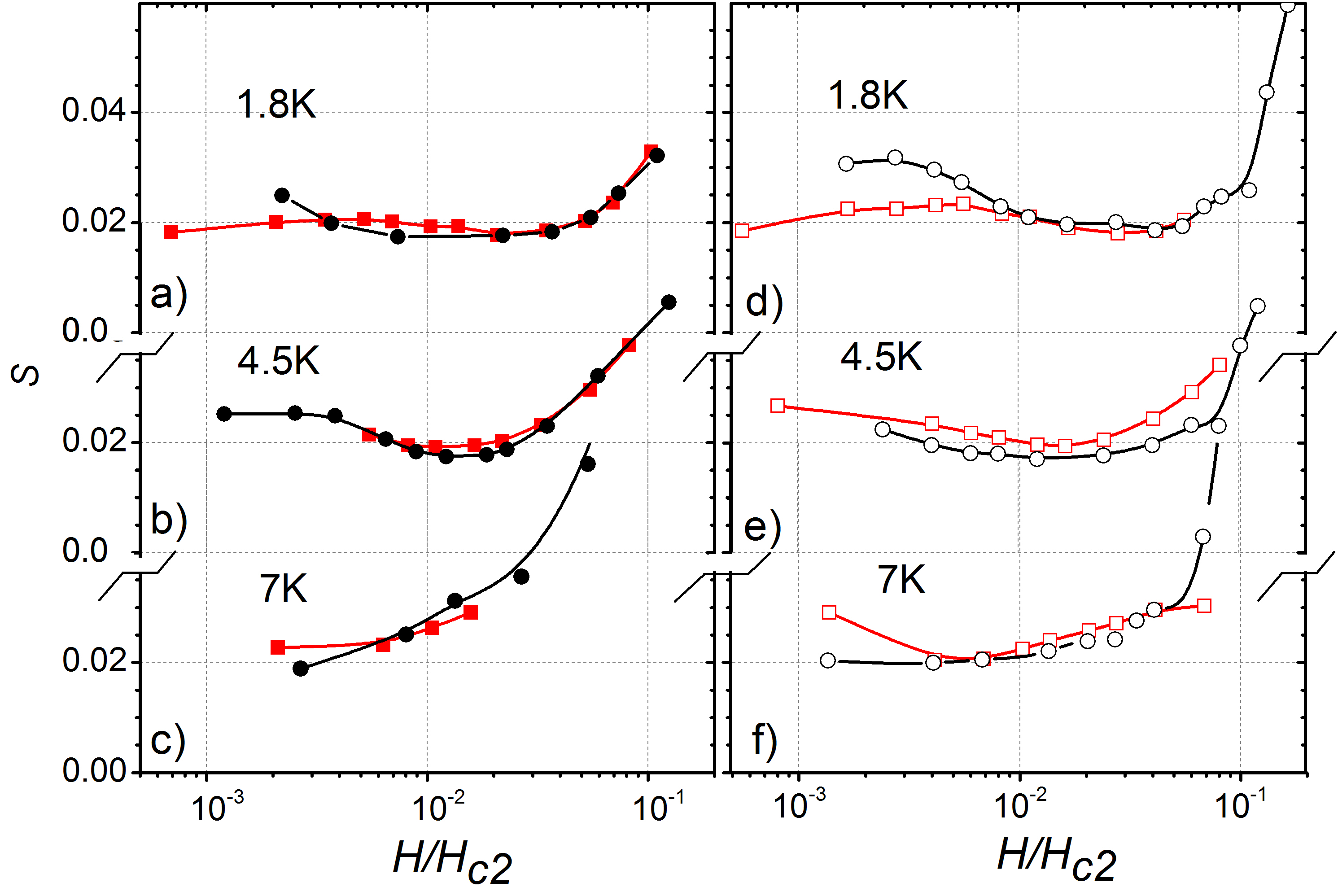}
\end{center}
\caption{Flux creep rates $(S)$ at a) 1.8\,K, b) 4.5\,K, and c) 7\,K in [f-SC] (circular black symbols) and [f-SC-irr] (squared red symbols) as function of the magnetic field normalized by $H_{c2}$. The $S$ values for the embedded sample are shown in  d), e) and f) panels also as a function of the field normalized by $H_{c2}$.}\label{S comp}
\end{figure}

Fig. \ref{S comp} presents the flux creep rate $S(H/H_{c2})$ for pristine and irradiated samples. 
Two main features are noticeable, the upturn at low fields (specially observed at low temperatures and in pristine samples) and the crossover to fast creep. 
The upturn at low fields is usually attributed to self-field effects \cite{PhysRevB.85.014522}.
However, and as we discuss in section \ref{3.4}, the large $S$ values at low temperatures and low fields are also related to single vortex pinning in a weak potential. 
In addition, the noticeable reduction observed for irradiated samples suggests a change in the flux creep mechanisms. 
On the other hand, the crossover to fast creep that appears shifted to lower fields in free-standing samples, remains unchanged after irradiation. Moreover, this crossover was reported to remain unchanged after proton irradiation in other FeBS \cite{PhysRevB.85.014522}, which indicates that it can be associated with an intrinsic increment of the thermal fluctuations of the system \cite{0953-2048-28-5-055011}. 
In addition, it appears to remain unchanged also for the embedded sample but the reduced field necessary for its occurrence is shifted to higher $H/H_{c2}$ and is sharper than the one observed in the free-standing samples.\cite{PhysRevB.85.014522} 

\subsection{Collective pinning energy and characteristics glassy exponents, $\mu$}\label{3.4}

Motivated by the differences of the vortex dynamics of pristine and proton irradiated samples, in this section we analyze the flux creep mechanism using the collective creep theory \cite{RevModPhys.66.1125}. 
This model considers that every single-vortex-line is pinned by the collective action of many weak point-like pinning centers. The pinning energy, $U$, results from a competition between the pinning potential and the elastic deformation of the vortices. 
At low magnetic fields, in the so-called single-vortex regime (SVR), the vortex-vortex interaction is negligible compared to the vortex-defect interaction. At higher fields, vortex-vortex interactions become dominant, and the vortices are collectively trapped as bundles. The normalized relaxation rate is given by 
\begin{equation}\label{eqS}
\ S=-\frac{\text{d}ln(J)}{\text{d}ln(t)}= \frac{T}{U_0+\mu T ln(t/t_0)}=\frac{T}{U}\Big(\frac{J}{J_c}\Big)^\mu,
\end{equation}   
where $\mu>$ 0 is the glassy exponent, $U_0$ and $t_0$ are characteristic energy and time scales, respectively. 
The activation energy as a function of the current density, $J$, in a glassy vortex phase is given by
\begin{equation}\label{}
\ U(J)=\frac{U_0(T)}{\mu} \Big[\Big(\frac{J_0}{ J}\Big)^{\mu} -1\Big].
\end{equation}
The glassy exponent $\mu$ depends on the dimension and length scales for the vortex lattice. 
According to the collective-pinning model, in the presence of random point defects and in the three-dimensional case, it results $\mu$ = 1/7 for SVR, 3/2 or 5/2 for small bundle ($sb$) and 7/9 for large-bundle ($lb$). 
Experimentally, the glassy exponents can be determined by the extended Maley's method \cite{PhysRevB.42.2639}. 
The time decay of $J$ is given by
\begin{equation}\label{}
 J=J_c [1+(\mu T/U_0) ln(t/t_0)]^{-1/\mu}.
\end{equation}
The effective activation energy $U_{eff}(J)$ can be obtained from experimental data considering the approximation in which the current density decays as
\begin{equation}\label{}
 \frac{\text{d}J}{\text{d}t}=-\frac{J_c}{T} e^{-U_{eff}(J)/T}.
\end{equation}
 The final equation for the pinning energy is
\begin{equation}\label{}
\ U_{eff}=-T\cdot [ln|dJ/dt|-C],
\end{equation}
 where $C=ln(J_c/T)$ is a nominally constant factor. 
For an overall analysis it is necessary to consider the function $G (T)$, which results in\cite{PhysRevB.46.3050}
\begin{equation}\label{}
\ U_{eff} (J,T=0) \sim U_{eff}  (J,T)/G(T).
\end{equation}

\begin{figure}[t]
\begin{center}
\includegraphics[width=0.46\textwidth]{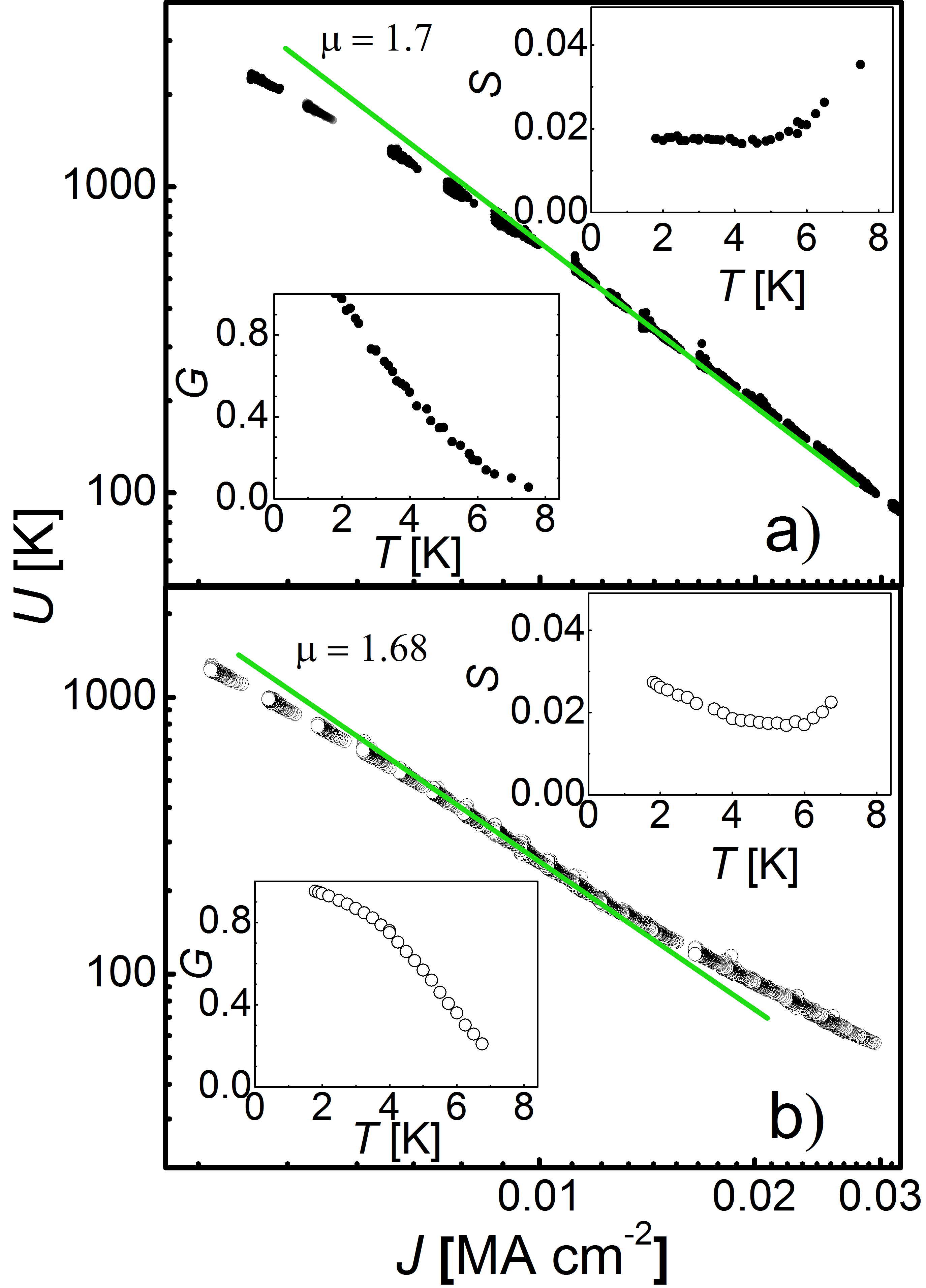}
\end{center}
\caption{Maley analysis with $\mu_0H = 0.1$\,T for the samples [f-SC] in a) and [e-SC] in b). $C = 13$ was used in both cases. $Inset:$ Temperature dependence of the creep relaxation rate $S(T)$ using $\mu_0H = 0.1$\,T (right) and $G(T)$ (left).}\label{Maley virgen}
\end{figure}

We performed $S(T)$ measurements at $\mu_0H = 0.1$\,T for the samples [f-SC], [e-SC], [f-SC-irr] and [e-SC-irr]. 
This field was selected because it is larger than the self-field (estimated as $\sim J_c \times d$) in all the samples. 
Figs. \ref{Maley virgen}a and \ref{Maley virgen}b show the results obtained for pristine samples and different configurations for the sample mounting. The insets present $S(T)$ (right) and $G(T)$ (left). 
Although there are remarkable differences in $S (T)$ at low temperatures (associated with different pinning mechanisms), both configurations display a plateau at intermediate temperatures (i.e. $\sim T_c/2$). 
At high temperatures the flux creep rates increase as a consequence of the expected thermal smearing of the pinning potential. 
In the limit of $J \ll J_c$, $\mu$ can be estimated as $\Delta ln U(J) / \Delta ln J$ \cite{PhysRevLett.78.3181}. 
At intermediate temperatures (in which $S(T)$ presents a plateau with $S\sim 0.018$) the slopes $\Delta ln U(J) / \Delta ln J$ are $\mu$ = 1.7 and  $\mu$ = 1.68 for the [f-SC] and [e-SC], respectively. These values are within the prediction for small bundles in random disorder\cite{RevModPhys.66.1125}. 
Similar values of $\mu$ can be expected at intermediate fields where $S (H) \sim constant$ (see Figs. \ref{JyS}a, \ref{JyS}b and \ref{JyS}c).  
Considering equation \ref{eqS}, the plateau of $S(T)$ is well described by  $S=1/(\mu ln(t/t_0) \sim 0.017$ (with $U_0 \ll  \mu T ln(t/t_0)$). 
Under this approximation and $\mu \sim$ 1.7, we obtained $ln(t/t_0)\sim$ 34, in good agreement with previously reported values for other FeBS\cite{0953-2048-27-9-095004, PhysRevB.86.094527}.

\begin{figure}[t]
\begin{center}
\includegraphics[width=0.46\textwidth]{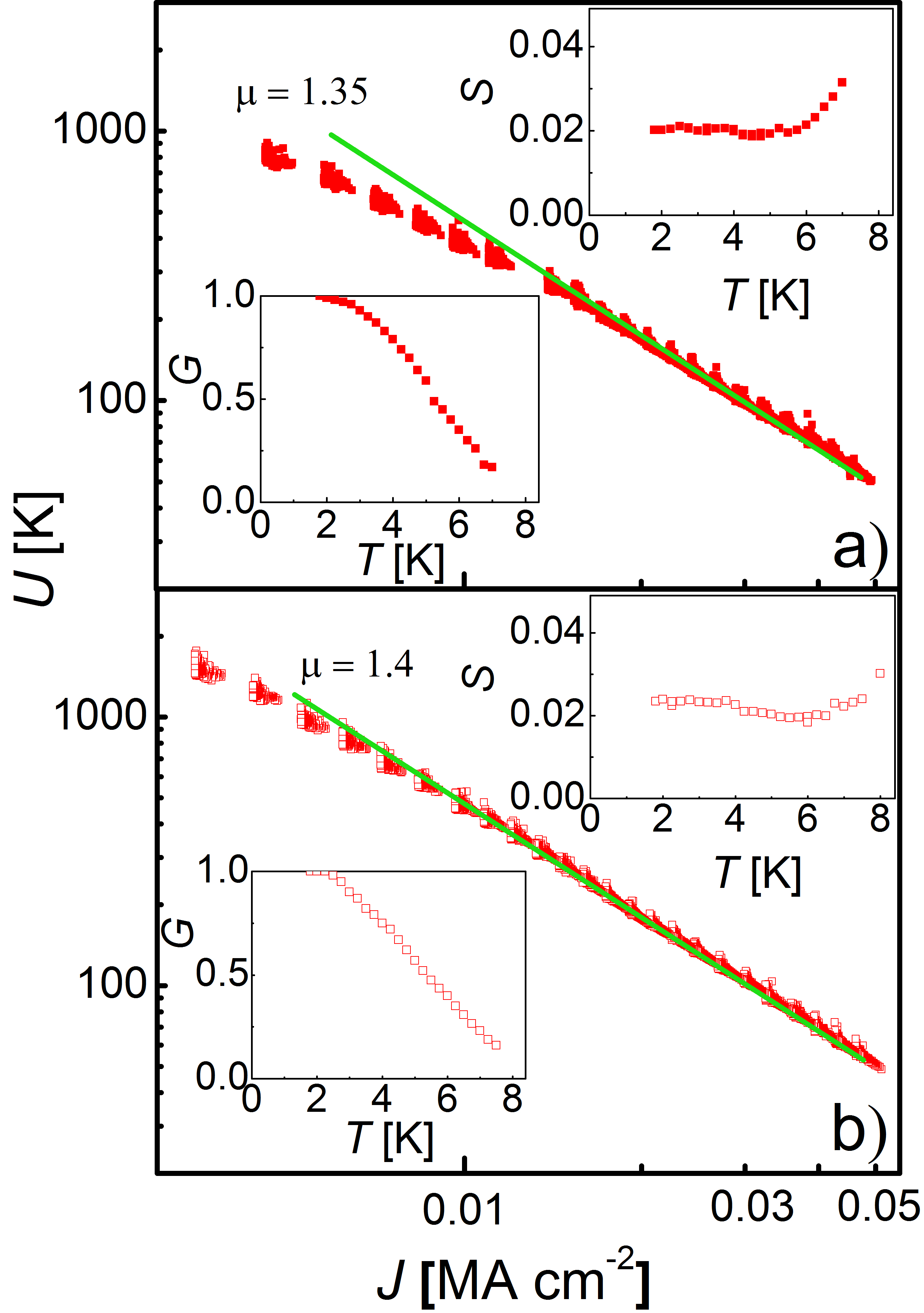}
\end{center}
\caption{Maley analysis with $\mu_0H = 0.1$\,T of the proton irradiated $\beta$-FeSe crystal with different mounting configurations, [f-SC-irr] in a) and [e-SC-irr] in b). $C = 13$ was used in both cases. $Inset:$ Temperature dependence of the creep relaxation rate $S(T)$ with $\mu_0H = 0.1$\,T (right) and $G(T)$ (left).}\label{Maley}
\end{figure}
Figs. \ref{Maley}a, \ref{Maley}b show the Maley analysis for irradiated samples, while the insets present $S(T)$ (right) and $G(T)$ (left). 
Both mounting configurations display $S(T) \sim 0.02$ at low and intermediate temperatures. It is noticeable that the large $S$ values observed in [e-SC] at low temperatures are suppressed by the irradiation. The glassy exponents obtained from $\Delta ln U(J) / \Delta ln J$ at intermediate temperatures are $\mu = 1.35$ and  $\mu = 1.4$ for [f-SC-irr] and [e-SC-irr], respectively. 
Since $S (T)$ is approximately constant from 1.8\,K to intermediate temperatures, the increment in $U_0$ due to the irradiation can be inferred to be very small. 
It is important to note that the reduction of $\mu$ at intermediate fields by proton irradiation is in agreement with other superconductors such as FeBS and cuprates\cite{0953-2048-28-5-055011, haberkorn2015influence}.

\begin{figure}[h]
\begin{center}
\includegraphics[width=0.46\textwidth]{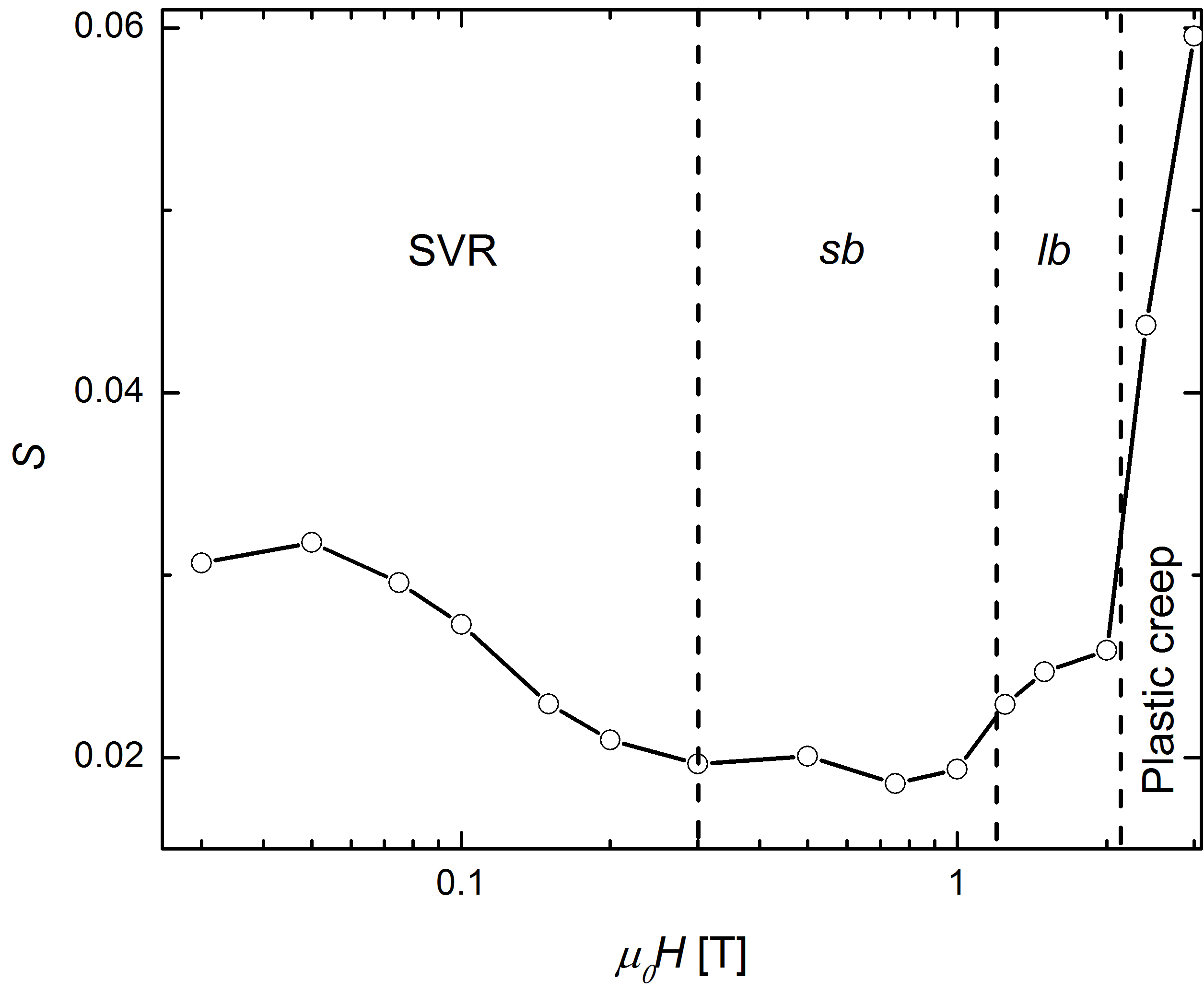}
\end{center}
\caption{Flux creep rate at 1.8\,K as function of field for the sample [e-SC]. The dotted lines indicate the crossovers between the different creep regimes.}\label{fig9}
\end{figure}

As we mentioned before, the vortex dynamics in [e-SC] is in agreement with the expectation for weak pinning produced by random point defects, which is considered for the collective creep theory \cite{RevModPhys.66.1125}. The characteristic glassy exponent $\mu$ theoretically predicted for random disorder is usually not observed experimentally, which can be related to mixed pinning landscapes (weak and strong pinning centers). In this sense, the embedded samples, in which the density of TBs is reduced, may allow a better observation of the different crossovers.
In the following, we analyze the vortex crossovers between SVR, $sb$ and $lb$ regimes present in [e-SC] and those predicted by the model (see Fig. \ref{fig9}). 
The SVR corresponds to weak fields where the distance between the vortex lines is large and their interaction is small compared to the interaction between the vortices and the quenched random potential of the defects. The SVR is typically not observed except under particular experimental conditions\cite{C2CE26857D}. 
This is partially because the range of magnetic fields where the SVR appears is strongly suppressed by temperature, and partially because it is usually masked by a low density of strong pinning centers (such TBs or nanoprecipitates). 
In addition, the values of $\mu$ experimentally determined are not always discrete, but rather present a gradual change\cite{PhysRevB.50.7188}.
The SVR occurs at low fields during the initial stage of the relaxation when  $J < J_c$. Single-vortex collective pinning is expected as long as $\gamma\cdot L_c < a_0$, where $L_c$ is the Larkin length, which can be calculated as $L_c = \gamma^{-1}\cdot\xi\cdot(J_0/J_c)^{1/2}$, and $a_0$ is the inter-vortex distance. For $\gamma L_c > a_0$, the interaction between vortices becomes important and the relaxation slows down. 
This regime is associated with relaxation by vortex bundles, and a new crossover from $\mu = 3/2$ (sb) to $\mu = 7/9$ (lb) is expected. 
The $U_c (H \parallel c)$ in the SVR can be estimated as\cite{RevModPhys.66.1125}
\begin{equation}\label{}
\ U_c^{SVR} \sim T_c [(J_c^c (1-T/T_c ))/(J_0 G_i)]^{(1/2)}.
\end{equation}
Considering $J_c (T = 1.8$\,K) $\sim 0.066$ MA$\cdot$ cm$^{-2}$ and $J_0 \sim 11.3$ MA$\cdot$ cm$^{-2}$, we obtained $U_c^{SVR} (0) \sim 90$\,K. 
This value is of the same order than the expectations considering the large $S$ values observed at the upturn at 1.8\,K. Using equation \ref{eqS} with $ln(t/t_0) \sim 34$ and $\mu = 1/7$, values of $S\sim 0.022 - 0.03$ are obtained for $U \sim 50 - 70$\,K. The crossover between SVR and $sb$ (in anisotropic superconductor with $H \parallel c$) is expected at $B_{sb}= \beta_{sb}(J_c/ J_0)H_{c2}$ , with $\beta_{sb}\sim 5$. 
According to this model, for [e-SC] $B_{sb} \sim 0.45$\,T, close to the experimental crossover field (0.2--0.3\,T) indicated as a dotted line in Fig. \ref{fig9}. 

In addition, the collective model also predicts the crossover from $sb$ to $lb$ when
$B_{lb} (0)\sim \beta_{lb} H_{c2} (J_{SV}/J_0 )( [ln⁡(\kappa^2  J_{SV}/J_0 )]^{(2/3)}$,
 with $\beta_{lb}=2$ \cite{PhysRevB.50.7188}. Using $\kappa= \lambda_{ab} (0)/ \xi_{ab} (0) \sim 100$, we obtained $B_{lb} (0) \sim 1.2$\,T, 
which agrees well with the second dotted line in Fig. \ref{fig9} placed at $\sim$1.1\,T. 
 Finally, the $lb$ regime disappears at high fields due to an increment in the vortex fluctuations and a crossover from elastic to plastic creep takes place \cite{PhysRevB.78.224506}.

\section{Conclusion}

In summary, we studied the vortex dynamics for $\beta$-FeSe single crystals. The results show that $T_c$ is affected by the mounting configuration method. Free-standing crystals present a superconducting critical temperature $T_c = 8.4(1)$\,K that increases to $10.5(1)$\,K when the crystals are fixed to the sample holder using GE-7031 varnish. In addition, we observe a remarkable influence of the mounting on the resulting $J_c$ and flux creep rates. 
The differences could be understood by considering the differences in the pinning landscape. 
For crystals held with GE-7031 varnish, the pinning may be mainly produced by random point defects whereas the free-standing sample presents a mix pinning landscape produced by random point defects and TBs. The irradiation with 3 MeV proton enhances $J_c$ and affect the vortex dynamics for both mounting configurations, with a larger effect on the embedded samples. 
From Maley analysis, we observe that the glassy exponent $\mu$ at intermediate temperatures and intermediate fields changes from $\sim 1.7$  to 1.35-1.4 after irradiation. 

\section*{Acknowledgements}
This work has been supported by Agencia Nacional de Promoci{\'o}n Cient{\'i}fica y Tecnol{\'o}gica PICT2015-2171 and PICT2014-1265; by Consejo Nacional de Investigaciones Cient{\'i}ficas y T\'ecnicas PIP2014-0164 and by Sectyp U.N.Cuyo 06/C504. N. H., G. N. and M. L. A. are members of the Instituto de Nanociencia y Nanotecnolog{\'i}a (Argentina).

\vspace {10 mm}

\bibliographystyle{iopart-num}
\bibliography{mibib}{}

\providecommand{\newblock}{}
\begin{thebibliography}{10}
\expandafter\ifx\csname url\endcsname\relax
  \def\url#1{{\tt #1}}\fi
\expandafter\ifx\csname urlprefix\endcsname\relax\def\urlprefix{URL }\fi
\providecommand{\eprint}[2][]{\url{#2}}

\bibitem{SUST0953-2048-25-8-084008}
Tamegai T, Taen T, Yagyuda H, Tsuchiya Y, Mohan S, Taniguchi T, Nakajima Y,
  Okayasu S, Sasase M, Kitamura H, Murakami T, Kambara T and Kanai Y 2012 {\em
  Superconductor Science and Technology\/} {\bf 25} 084008
  \urlprefix\url{http://stacks.iop.org/0953-2048/25/i=8/a=084008}

\bibitem{PhysRevB.78.224506}
Prozorov R, Ni N, Tanatar M~A, Kogan V~G, Gordon R~T, Martin C, Blomberg E~C,
  Prommapan P, Yan J~Q, Bud'ko S~L and Canfield P~C 2008 {\em Phys. Rev. B\/}
  {\bf 78}(22) 224506
  \urlprefix\url{http://link.aps.org/doi/10.1103/PhysRevB.78.224506}

\bibitem{RevModPhys.66.1125}
Blatter G, Feigel'man M~V, Geshkenbein V~B, Larkin A~I and Vinokur V~M 1994
  {\em Rev. Mod. Phys.\/} {\bf 66}(4) 1125--1388
  \urlprefix\url{http://link.aps.org/doi/10.1103/RevModPhys.66.1125}

\bibitem{PhysRevLett.115.027006}
Watson M~D, Yamashita T, Kasahara S, Knafo W, Nardone M, B\'eard J, Hardy F,
  McCollam A, Narayanan A, Blake S~F, Wolf T, Haghighirad A~A, Meingast C,
  Schofield A~J, v~L\"ohneysen H, Matsuda Y, Coldea A~I and Shibauchi T 2015
  {\em Phys. Rev. Lett.\/} {\bf 115}(2) 027006
  \urlprefix\url{http://link.aps.org/doi/10.1103/PhysRevLett.115.027006}

\bibitem{LouLT27}
Amig{\'o} M~L, Crivillero V~A, Franco D~G and Nieva G 2014 {\em Journal of
  Physics: Conference Series\/} {\bf 568} 022005
  \urlprefix\url{http://stacks.iop.org/1742-6596/568/i=2/a=022005}

\bibitem{doi:10.1143/JPSJ.78.074712}
Mizuguchi Y, Tomioka F, Tsuda S, Yamaguchi T and Takano Y 2009 {\em Journal of
  the Physical Society of Japan\/} {\bf 78} 074712 (\textit{Preprint}
  \eprint{http://dx.doi.org/10.1143/JPSJ.78.074712})
  \urlprefix\url{http://dx.doi.org/10.1143/JPSJ.78.074712}

\bibitem{PhysRevB.91.174510}
Kn\"oner S, Zielke D, K\"ohler S, Wolf B, Wolf T, Wang L, B\"ohmer A, Meingast
  C and Lang M 2015 {\em Phys. Rev. B\/} {\bf 91}(17) 174510
  \urlprefix\url{http://link.aps.org/doi/10.1103/PhysRevB.91.174510}

\bibitem{doi:10.1063/1.4903922}
Wang W, Li J, Yang J, Gu C, Chen X, Zhang Z, Zhu X, Lu W, Wang H~B, Wu P~H,
  Yang Z, Tian M, Zhang Y and Moshchalkov V~V 2014 {\em Applied Physics
  Letters\/} {\bf 105} 232602 (\textit{Preprint}
  \eprint{http://dx.doi.org/10.1063/1.4903922})
  \urlprefix\url{http://dx.doi.org/10.1063/1.4903922}

\bibitem{PhysRevB.94.064521}
Teknowijoyo S, Cho K, Tanatar M~A, Gonzales J, B\"ohmer A~E, Cavani O, Mishra
  V, Hirschfeld P~J, Bud'ko S~L, Canfield P~C and Prozorov R 2016 {\em Phys.
  Rev. B\/} {\bf 94}(6) 064521
  \urlprefix\url{http://link.aps.org/doi/10.1103/PhysRevB.94.064521}

\bibitem{PhysRevLett.103.057002}
McQueen T~M, Williams A~J, Stephens P~W, Tao J, Zhu Y, Ksenofontov V, Casper F,
  Felser C and Cava R~J 2009 {\em Phys. Rev. Lett.\/} {\bf 103}(5) 057002
  \urlprefix\url{http://link.aps.org/doi/10.1103/PhysRevLett.103.057002}

\bibitem{PhysRevB.79.180508}
Tanatar M~A, Kreyssig A, Nandi S, Ni N, Bud'ko S~L, Canfield P~C, Goldman A~I
  and Prozorov R 2009 {\em Phys. Rev. B\/} {\bf 79}(18) 180508
  \urlprefix\url{http://link.aps.org/doi/10.1103/PhysRevB.79.180508}

\bibitem{PhysRevLett.109.137004}
Song C~L, Wang Y~L, Jiang Y~P, Wang L, He K, Chen X, Hoffman J~E, Ma X~C and
  Xue Q~K 2012 {\em Phys. Rev. Lett.\/} {\bf 109}(13) 137004
  \urlprefix\url{http://link.aps.org/doi/10.1103/PhysRevLett.109.137004}

\bibitem{nature}
Maggio-Aprile I, Renner C, Erb A, Walker E and Fischer O 1997 {\em Nature\/}
  {\bf 390} 487--490 \urlprefix\url{http://dx.doi.org/10.1038/37312}

\bibitem{PhysRevB.92.144509}
Sun Y, Pyon S, Tamegai T, Kobayashi R, Watashige T, Kasahara S, Matsuda Y and
  Shibauchi T 2015 {\em Phys. Rev. B\/} {\bf 92}(14) 144509
  \urlprefix\url{http://link.aps.org/doi/10.1103/PhysRevB.92.144509}

\bibitem{PhysRevB.88.174512}
Abdel-Hafiez M, Ge J, Vasiliev A~N, Chareev D~A, Van~de Vondel J, Moshchalkov
  V~V and Silhanek A~V 2013 {\em Phys. Rev. B\/} {\bf 88}(17) 174512
  \urlprefix\url{http://link.aps.org/doi/10.1103/PhysRevB.88.174512}

\bibitem{Amigó2015}
Amig{\'o} M~L, Ale~Crivillero M~V, Franco D~G, Guimpel J and Nieva G 2015 {\em
  Journal of Low Temperature Physics\/} {\bf 179} 15--20 ISSN 1573-7357
  \urlprefix\url{http://dx.doi.org/10.1007/s10909-014-1255-9}

\bibitem{Morelock:ks5362}
Morelock C~R, Suchomel M~R and Wilkinson A~P 2013 {\em Journal of Applied
  Crystallography\/} {\bf 46} 823--825
  \urlprefix\url{https://doi.org/10.1107/S0021889813005955}

\bibitem{PhysRevB.42.2639}
Maley M~P, Willis J~O, Lessure H and McHenry M~E 1990 {\em Phys. Rev. B\/} {\bf
  42}(4) 2639--2642
  \urlprefix\url{http://link.aps.org/doi/10.1103/PhysRevB.42.2639}

\bibitem{RevModPhys.68.911}
Yeshurun Y, Malozemoff A~P and Shaulov A 1996 {\em Rev. Mod. Phys.\/} {\bf
  68}(3) 911--949
  \urlprefix\url{http://link.aps.org/doi/10.1103/RevModPhys.68.911}

\bibitem{WANG20171}
Wang X, Zhang Z, Wang W, Zhou Y, Kan X, Chen X, Gu C, Zhang L, Pi L, Yang Z and
  Zhang Y 2017 {\em Physica C: Superconductivity and its Applications\/} {\bf
  537} 1 -- 4 ISSN 0921-4534
  \urlprefix\url{http://www.sciencedirect.com/science/article/pii/S0921453416302581}

\bibitem{PhysRevLett.65.1164}
Civale L, Marwick A~D, McElfresh M~W, Worthington T~K, Malozemoff A~P,
  Holtzberg F~H, Thompson J~R and Kirk M~A 1990 {\em Phys. Rev. Lett.\/} {\bf
  65}(9) 1164--1167
  \urlprefix\url{http://link.aps.org/doi/10.1103/PhysRevLett.65.1164}

\bibitem{PhysRevB.85.014522}
Haberkorn N, Maiorov B, Usov I~O, Weigand M, Hirata W, Miyasaka S, Tajima S,
  Chikumoto N, Tanabe K and Civale L 2012 {\em Phys. Rev. B\/} {\bf 85}(1)
  014522 \urlprefix\url{http://link.aps.org/doi/10.1103/PhysRevB.85.014522}

\bibitem{0953-2048-27-9-095004}
Haberkorn N, Kim J, Maiorov B, Usov I, Chen G~F, Yu W and Civale L 2014 {\em
  Superconductor Science and Technology\/} {\bf 27} 095004
  \urlprefix\url{http://stacks.iop.org/0953-2048/27/i=9/a=095004}

\bibitem{C2CE26857D}
Chareev D, Osadchii E, Kuzmicheva T, Lin J~Y, Kuzmichev S, Volkova O and
  Vasiliev A 2013 {\em CrystEngComm\/} {\bf 15}(10) 1989--1993
  \urlprefix\url{http://dx.doi.org/10.1039/C2CE26857D}

\bibitem{PhysRevB.50.7188}
Civale L, Krusin-Elbaum L, Thompson J~R and Holtzberg F 1994 {\em Phys. Rev.
  B\/} {\bf 50}(10) 7188--7191
  \urlprefix\url{https://link.aps.org/doi/10.1103/PhysRevB.50.7188}

\bibitem{PhysRevB.81.174517}
van~der Beek C~J, Rizza G, Konczykowski M, Fertey P, Monnet I, Klein T, Okazaki
  R, Ishikado M, Kito H, Iyo A, Eisaki H, Shamoto S, Tillman M~E, Bud'ko S~L,
  Canfield P~C, Shibauchi T and Matsuda Y 2010 {\em Phys. Rev. B\/} {\bf
  81}(17) 174517
  \urlprefix\url{http://link.aps.org/doi/10.1103/PhysRevB.81.174517}

\bibitem{doi:10.1063/1.4821440}
Jia Y, LeRoux M, Miller D~J, Wen J~G, Kwok W~K, Welp U, Rupich M~W, Li X,
  Sathyamurthy S, Fleshler S, Malozemoff A~P, Kayani A, Ayala-Valenzuela O and
  Civale L 2013 {\em Applied Physics Letters\/} {\bf 103} 122601
  (\textit{Preprint} \eprint{http://dx.doi.org/10.1063/1.4821440})
  \urlprefix\url{http://dx.doi.org/10.1063/1.4821440}

\bibitem{0953-2048-28-5-055011}
Haberkorn N, Kim J, Gofryk K, Ronning F, Sefat A~S, Fang L, Welp U, Kwok W~K
  and Civale L 2015 {\em Superconductor Science and Technology\/} {\bf 28}
  055011 \urlprefix\url{http://stacks.iop.org/0953-2048/28/i=5/a=055011}

\bibitem{PhysRevB.46.3050}
Ossandon J~G, Thompson J~R, Christen D~K, Sales B~C, Sun Y and Lay K~W 1992
  {\em Phys. Rev. B\/} {\bf 46}(5) 3050--3058
  \urlprefix\url{http://link.aps.org/doi/10.1103/PhysRevB.46.3050}

\bibitem{PhysRevLett.78.3181}
Thompson J~R, Krusin-Elbaum L, Civale L, Blatter G and Feild C 1997 {\em Phys.
  Rev. Lett.\/} {\bf 78}(16) 3181--3184
  \urlprefix\url{https://link.aps.org/doi/10.1103/PhysRevLett.78.3181}

\bibitem{PhysRevB.86.094527}
Taen T, Nakajima Y, Tamegai T and Kitamura H 2012 {\em Phys. Rev. B\/} {\bf
  86}(9) 094527
  \urlprefix\url{http://link.aps.org/doi/10.1103/PhysRevB.86.094527}

\bibitem{haberkorn2015influence}
Haberkorn N, Kim J, Su{\'a}rez S, Lee J~H and Moon S 2015 {\em Superconductor
  Science and Technology\/} {\bf 28} 125007

\end{thebibliography}

\end{document}